\documentclass[
aps,pra,
twocolumn,
showpacs,
superscriptaddress,
floatfix,
]{revtex4-1}
\usepackage[dvipdfmx]{graphicx}
\usepackage{dcolumn}
\usepackage{bm}
\usepackage{ulem}
\usepackage{amsmath}
\usepackage{amssymb}
\usepackage{txfonts}
\usepackage{hyperref}
\usepackage{color} 
\hypersetup{
  colorlinks=true,
  linkcolor=[rgb]{0.85,0.00,0.15},
  citecolor=[rgb]{0.0,0.15,0.95},
  urlcolor=[rgb]{0.0,0.15,0.95},
  setpagesize=false
}

\begin{document}  

\title{
Electronic properties and applications of MXenes: a theoretical review
}

\author{Mohammad Khazaei
\footnote{E-mail address: khazaei@riken.jp}}
\author{Ahmad Ranjbar}
\affiliation{Computational Materials Science Research Team, RIKEN Advanced Institute for Computational Science (AICS), Kobe, Hyogo 650-0047, Japan}

\author{Masao Arai}
\author{Taizo Sasaki}
\affiliation{International Center for Materials Nanoarchitectonics, National Institute for Materials Science (NIMS), 1-1 Namiki, Tsukuba 305-0044, Ibaraki, Japan}

\author{Seiji Yunoki}
\affiliation{Computational Materials Science Research Team, RIKEN Advanced Institute for Computational Science (AICS), Kobe, Hyogo 650-0047, Japan}
\affiliation{Computational Condensed Matter Physics Laboratory, RIKEN, Wako, Saitama 351-0198, Japan}
\affiliation{Computational Quantum Matter Research Team, RIKEN Center for Emergent Matter Science (CEMS), Wako, Saitama 351-0198, Japan}

\date{\today}

\begin{abstract}
 Recent chemical exfoliation of layered MAX phase compounds 
to novel two-dimensional transition metal carbides and nitrides, so called MXenes, has brought new opportunity to materials 
science and technology. This review highlights the computational attempts that have been made to understand the physics 
and chemistry of this very promising family of advanced two-dimensional materials, and to exploit their novel and exceptional 
properties for electronic and energy harvesting applications.
\end{abstract}

\maketitle
\section{Introduction}

It has been demonstrated in recent extensive studies that low-dimensional systems containing transition metals 
with open $d$-orbital shells may exhibit a multitude 
of interesting properties because of different oxidation and spin states, and relatively large spin-orbit coupling 
of the transitions metals. 
Hence, transition metal-based low-dimensional systems provide an excellent ground for exploring and exploiting the internal degrees of 
freedom of electrons $-$ charge, orbital, and spin $-$ and their interplay for fundamental research and device applications~\cite{G.Fiori2014,S.Z.Butler2013,Q.H.Wang2012,G.R.Bhimanapati2015}. 
There are many transition metal-based low dimensional systems in the literature, e.g. dichalcogenides, which have been or may potentially be exfoliated in 
the experiments~\cite{J.N.Coleman2011,H.Zhang2015}. Among them, nowadays, 
MXenes~\cite{M.Naguib2011,M.Naguib2012} are truly at the cutting 
edge of materials research and promise new scientific and technological 
horizons.

MXenes~\cite{M.Naguib2011,M.Naguib2012} are a new class of two-dimensional (2D) transition metal carbides and nitrides with chemical formula of M$_{n+1}$X$_n$ 
(M= Sc, Ti, V, Cr, Zr, Nb, Mo, Hf, Ta; X= C, N; $n$=1--3) that have recently been synthesized through etching 
MAX phases~\cite{M.Khazaei2014_1,M.W.Barsoum2000,Z.M.Sun2011,M.F.Cover2009,M.Ashton2016_1,M.Khazaei2014_2}. 
These 2D systems have been named 
as MXenes because they originate from the MAX phases by removing ``A'' elements and because they are 
structurally analogous to graphene~\cite{M.Naguib2011,M.Naguib2012}. Very recently, 
significant progress in the growth of high quality crystalline MXenes has been achieved by the chemical 
vapor deposition technique~\cite{C.Xu2015,D.Geng2017}. 
Moreover, the family of MXenes has been lately expanded to ordered double transition metals carbides 
M$'_2$M$''$C$_2$ and M$'_2$M$''_2$C$_3$~\cite{B.Anasori2015}. 
Considering the large number of compositional variety of the MAX phase compounds, a large number of 2D 
MXenes with unprecedented properties is expected to be produced. 

Experimentally, MXenes have already found applications as 
transparent conductors~\cite{J.Halim2014,M.Mariano2016,Y.Yang2017}, 
field effect transistors~\cite{S.Lai2015},
supercapacitors~\cite{M.R.Lukatskaya2013,M.Chidiu2014,R.B.Rakhi2015}, 
Li ion batteries~\cite{M.Naguib2013,Y.Xie2014_1}, 
electromagnetic interface shielders~\cite{F.Shahzad2016}, fillers in polymeric composites~\cite{X.Zhang2013}, hybrid nanocompositites~\cite{M.Xue2017},  
purifiers~\cite{Q.Peng2014,J.Guo2015}, 
dual-responsive surfaces~\cite{J.Chen2015}, 
suitable substrates for dyes~\cite{O.Mashtalir2014}, catalysts~\cite{G.Fan2017,Z.W.Seh2016}, promising materials for methane storage~\cite{F.Liu2016}, and photocatalysts for hydrogen production~\cite{J.Ran2017}, as well as being ceramic biomaterials with high photothermal conversion efficiency for cancer therapy \cite{H.Lin2016}. 
Theoretically, many applications have been proposed for MXenes in
electronic~\cite{M.Khazaei2013,A.N.Enyashin2013,Y.Xie2013,M.Khazaei2016_1,L.Feng2017}, 
magnetic \cite{C.Si2015,J.He2016,M.Je2016,G.Gao2016,J.Yang2016_1}, optical~\cite{H.Lashgari2014,Y.Bai2016}, 
thermoelectric~\cite{M.Khazaei2014,A.N.Gand2016,S.Kumar2016,X.H.Zha2016_1,X.H.Zha2016_2,X.H.Zha2016_3}, 
and sensing devices~\cite{X.F.Yu2015_1}, 
as well as being new potential materials for catalytic and photocatalytic reactions~\cite{C.Ling2016_1,L.Y.Gan2013_1,C.Ling2016_2,G.Gao2016_2,Z.Guo2016,H.Zhang2016}, hydrogen storage media~\cite{Q.Hu2013,Q.Hu2014}, and nanoscale superconductivity\cite{J.J.Zhang2017}. 
Some of MXenes are predicted to be topological insulators with large band gaps 
involving only $d$ orbitals~\cite{M.Khazaei2016_2,H.Weng2015_1,L.Li2016,C.Si2016,C.Si2016_2}. 
MXenes are also expected to be used as ultralow work function materials~\cite{M.Khazaei2015} and 
Schottky barrier junctions~\cite{L.Y.Gan2013_2,H.Zhao2015,Y.Lee2015,Y.Liu2016}.

As the experimental and theoretical studies described above have already suggested, 
MXenes are appealing 2D systems because of the following reasons. 
1) Owing to their ceramic nature, MXenes are chemically and mechanically stable. 
2) MXenes can be found in different forms of monolayer, few layers, and multilayers. 
3) MXenes can be synthesized as complex materials made of mixture 
of light and heavy transition metals, which makes it possible to tune the number of valence electrons 
and the relativistic spin-orbit coupling (SOC). 
This enhances their electronic functionality and mechanical stability. 
4) The thickness of MXene monolayers is controllable. 
This allows us to examine the quantum confinement related phenomena. 
5) The surfaces of MXenes can be functionalized with various 
chemical groups, which offers possibilities for surface state engineering. 
6) Similar to graphene, some of MXenes exhibit massless Dirac dispersions 
in their band structures near the Fermi 
level~\cite{M.Khazaei2016_2,H.Weng2015_1,L.Li2016,C.Si2016,H.Fashandi2015}. 
This opens broad possibilities for 
Dirac-based physics and applications~\cite{M.Khazaei2016_2,H.Weng2015_1,L.Li2016,C.Si2016}. 
These properties make MXenes unique among other known 2D materials although some 
of them have not been experimentally observed yet.

This article reviews the current status of theoretical studies on MXenes, highlighting 
the recent progress based on first-principles calculations. 
It updates and comprehends the previous excellent theoretical 
reviews in Refs.~\cite{A.L.Ivanovskii2013,Q.Tang2015} 
by providing insights 
into many physical and chemical characteristics and applications of MXenes.
For the experimental progresses on MXenes,
one can find comprehensive reviews in Refs.~\cite{M.Naguib2014,M.Naguib2015,J.C.Lei2015,V.Ng2016,B.Anasori2017_1}.  
The rest of this review is organized as follows. 
Sec.~\ref{sec:method} summarizes computational methods employed to obtain 
the theoretical results discussed in this review. 
Sec.~\ref{sec:max} briefly discusses the structural, 
mechanical, and electronic properties of the parent structures of MXenes, \textit{i.e.}, MAX phases.
Sec.~\ref{sec:mxene} explains first why the surfaces of 2D MXenes are usually saturated with 
chemical groups. Then, the effects of surface functionalization on mechanical, electronic, 
magnetic, optical, transport, and surfaces states of MXenes are discussed. 
Some of the proposed
applications of 2D MXenes as low work function electron emitters, 
catalyst for hydrogen evolution, promising thermoelectric materials, and  
energy storage media are also described. 
Sec.~\ref{sec:nano} summarizes the electronic structures of MXenes in the form of nanoribbons, 
nanotubes and heterostructures. Finally, outlook for future research on MXenes is 
provided in Sec.~\ref{sec:outlook}.

\section{Computational methods}\label{sec:method}

Before discussing the properties and applications of MXenes, it is worth summarizing briefly 
computational methods employed in the following theoretical studies of this review. 
Among different methods, density functional theory (DFT) has 
proven to be a reliable method to predict various physical and chemical phenomena in atomic scale. Hence, 
DFT calculations are performed to examine many properties and applications of MXenes. 
However, it is known that the simple DFT method may fail in predicting the values of band gap and van der Walls 
interaction, and properties of strongly correlated materials. 
For instance, the Perdew-Burke-Ernzerhof (PBE) version of the generalized gradient approximation (GGA) 
for the exchange-correlation functional~\cite{J.P.Perdew1996}, which is most often used in this review, 
underestimates the band gap. Hence, in some of the studies discussed below, the hybrid functional [Heyd-Scuseria-Ernzerhof (HSE06)]~\cite{J.Heyd2006,S.Baroni2001} is employed to improve the estimation of the band gap, 
and the semi-empirical DFT-D2 method (Grimme method)~\cite{S.J.Grimme2006} is used to 
treat properly the Van der Walls interaction. Furthermore, the DFT+U method~\cite{A.I.Liechtenstein1995,S.L.Dudarev1998}
is applied to correctly obtain 
magnetic order when the electron correlation in transition metals is important. 
Because of the approximate nature of the DFT-based methods~\cite{A.J.Cohen2008}, this review mainly 
focuses on the general trends, which are expected in the physics and chemistry of MXenes, 
rather than emphasizing quantitatively the exact values of various properties. 
In the following, the DFT-PBE method is employed for the theoretical studies, unless otherwise 
stated.  

\section{MAX phases}\label{sec:max}

\begin{figure}[t]
\centering
  \includegraphics[scale=0.222]{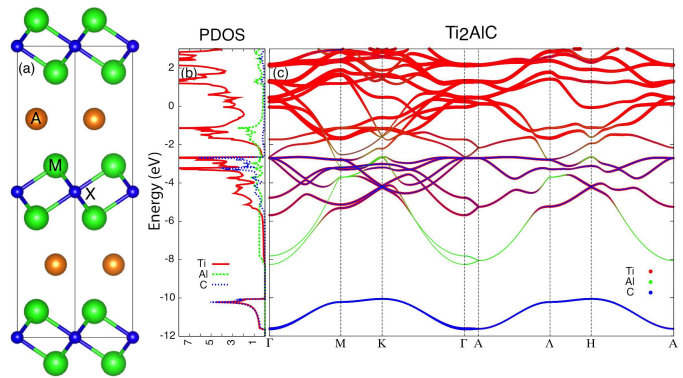}
  \caption{(a) Typical structure of M$_2$AX MAX phases. (b) Projected density of states 
  [PDOS (states/eV/cell)] and (c) projected band structure of Ti$_2$AlC~\cite{M.Khazaei2014_1}. 
  For clarity, the PDOS for Al is scaled by factor 3. 
  The Fermi energy is located at zero energy. 
  $\Gamma$(0,0,0), M(1/2,0,0), K(1/3,1/3,0), A(0,0,1/2), $\Lambda$(1/2,0,1/2), and
H(1/3,1/3,1/2) are high symmetric momenta in the Brillouin zone of Ti$_2$AlC.}
  \label{fig:max}
\end{figure}

MAX phases are a large family of hexagonal layered ceramic compounds with symmetry group of P6$_3$/mmc and with chemical formula of M$_{n+1}$AX$_n$ ($n$=1--3), 
where ``M'', ``A'', and ``X'' represent an early transition metal (Sc, Ti, V, Cr, Zr, Nb, Mo, Hf, and Ta), 
an element from groups 13--16 (Al, Si, P, S, Ga, Ge, As, In, and Sn) in the 
periodic table, and carbon and/or nitrogen, respectively~\cite{M.Khazaei2014_1,M.W.Barsoum2000,Z.M.Sun2011,M.F.Cover2009,M.Ashton2016_1,M.Khazaei2014_2}. 
Figure~\ref{fig:max}(a) shows
the typical crystal structure of M$_2$AX MAX phase compunds. 
Over 60 different MAX phase compounds have already been synthesized 
experimentally~\cite{M.W.Barsoum2000,Z.M.Sun2011}. 
MAX phases typically exhibit favorable properties of ceramics such as high structural stiffness, and also 
favorable properties of metals such as good electrical and thermal conductivity~\cite{M.Khazaei2014_1}. 
MAX phases have potential applications such as wear and corrosion protection, heat exchangers, lubricants, 
nozzles, and kiln furniture~\cite{M.W.Barsoum2000,Z.M.Sun2011}.

Based on first-principles calculations, elastic constants of more than 240 different M$_2$AC and M$_2$AN 
compounds with various M and A elements have been theoretically studied so far~\cite{M.F.Cover2009}. 
Among these, many of the compounds are found to satisfy the mechanical stability criteria for hexagonal 
structures: $C_{11}> |C_{12}|, C_{66}>0$, and $(C_{11}+C_{12})C_{33}-2C_{13}^2>0$, where 
$C_{ij}$ is the second rank tensor of elastic constant~\cite{J.F.Nye2010}. 
Therefore, many of them are elastically stable and have a great chance to be realized  
under appropriate experimental conditions. 
Another computational screening study has analyzed the formation energies of 216 pure M$_2$AX compounds and 
10314 solid solution (MM$'$)$_2$(AA$'$)(XX$'$)~\cite{M.Ashton2016_1}, and  
found that 3140 compounds, including 49 experimentally known M$_2$AX phases, 
exhibit formation energies of less than $-30$ meV/atom. 
A significant subsets of 301 compositions with formation energy lower than $-100$ meV/atom 
exhibit favorable synthesis conditions.

Recently, the family of MAX phases has also been theoretically extended to M$_2$AlB 
compounds~\cite{M.Khazaei2014_1}. 
The formation energy ($E_{\rm f}$) calculations find that 
E$_{\rm f}$(M$_2$AlN) < E$_{\rm f}$(M$_2$AlC) < E$_{\rm f}$(M$_2$AlB)~\cite{M.Khazaei2014_1}. 
This indicates that the M$_2$AlN phases will likely be obtained with higher purity in 
experiments. This trend of the formation energies for MAX phases is understood by considering several 
entangled factors, i.e., 
1) the strength of hybridization between $d$ orbitals of M and others ($p$ orbitals of Al, 
and $s$ and $p$ orbitals of X), 
2) the number of valence electrons that fill bonding, nonbonding, and antibonding states, and 3) the 
required energy for volume expansion of the transition metals to host the X elements~\cite{M.Khazaei2014_1}.

The first-principles calculations find that M$_2$AlB, M$_2$AlC, and M$_2$AlN are all 
metallic~\cite{M.Khazaei2014_1}. 
As shown in Figs.~\ref{fig:max}(b) and \ref{fig:max}(c) for Ti$_2$AlC, the states around 
the Fermi energy of these compounds are mainly dominated 
by $d$ orbitals of the transition metals. 
The common features of hybridization are evident in these compounds, where bonding (antibonding) states 
between $d$ orbitals of M and $p$ orbitals of Al or $p$ orbitals of X are located below (above) the 
Fermi energy. Between these bonding and antibonding states, nonbonding states of M are located near the Fermi energy. More detailed analysis shows that the 
number of valence electrons per unit cell (valence electron concentration), 
which fill the bonding and antiboding states, can affect and control the formation 
energy and the mechanical property of MAX phases significantly~\cite{M.Khazaei2014_1}.

\begin{figure}[t]
\centering
  \includegraphics[scale=0.23]{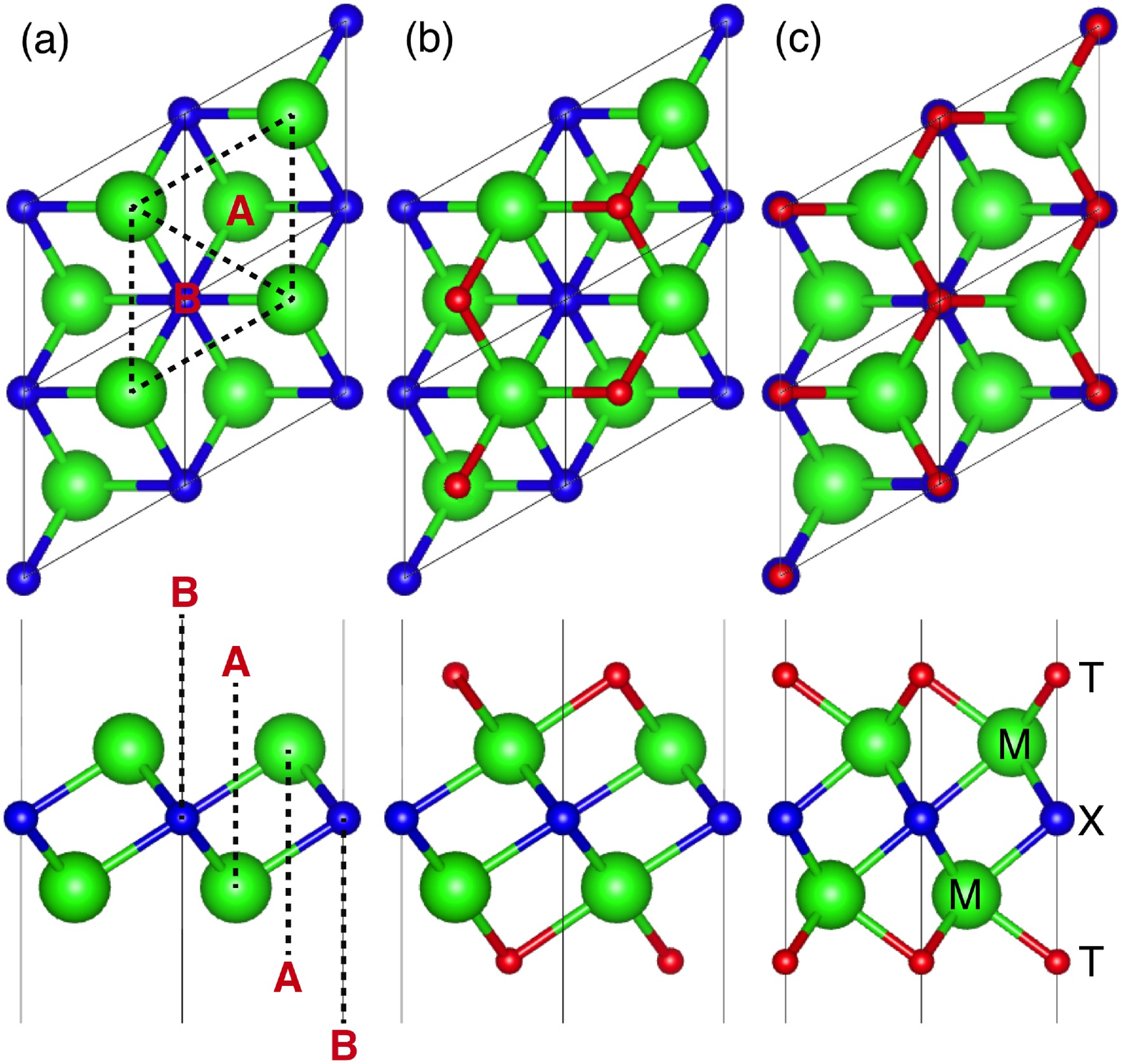}
  \caption{Top and side views of (a) pristine MXene M$_2$X. 
  Top and side views of (b) model 2 and (c) model 4 for functionalized 
  MXene (see the text). A and B indicate different types of hollow sites in (a). 
  M, X, and T denote transition metal (green), C/N (blue), and attached chemical groups such as F, O, and OH 
  (red), respectively.
  }
  \label{fig:mxene1}
\end{figure}

\section{2D MXenes}\label{sec:mxene}

2D MXenes are experimentally obtained when the ``A'' element is washed out from the MAX phases 
upon applying appropriate acid solution such as HF~\cite{M.Naguib2011,M.Naguib2012}. 
This is possible because in some of the MAX phases 
the bonding between elements A and M is weaker than those between elements M and X. 
However, it is experimentally observed that the outer layers are often saturated 
with F, O, and/or OH groups when MXenes are chemically exfoliated by HF acid solutions to eliminate 
the  ``A'' element~\cite{M.A.Hope2016,H.W.Wang2016,G.Sharma2016,D.Magne2016,K.D.Fredrickson2016}.

The exfoliation process of Ti$_3$AlC$_2$ to Ti$_3$C$_2$ has been 
theoretically studied in the presence of water and HF acid using \textit{ab initio} molecular dynamics 
simulations~\cite{P.Srivastava2016,A.Mishra2016}. There, it is shown 
that the spontaneous dissociation of HF and subsequent termination with H or F
at the edge of Ti atoms make Al-M bonding weak, consequently opening the interlayer gap. 
The further insertion of HF is facilitated  through the interlayer gap. This leads to the formation of 
AlF$_3$ and H$_2$, which eventually come out 
of the MAX phase. After these processes, the fluorinated MXene leaves 
behind~\cite{P.Srivastava2016,A.Mishra2016}. 

\begin{figure}[t]
\centering
  \includegraphics[scale=0.215]{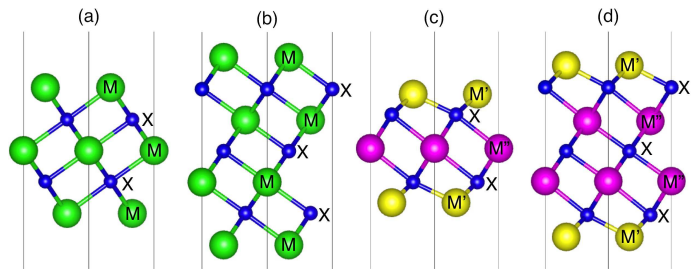}
  \caption{Side views of pristine (a) M$_3$X$_2$, (b) M$_4$X$_3$, (c) M$'_2$M$''$X$_2$, and (d) 
  M$'_2$M$''_2$X$_3$ MXenes, where M, M$'$, and M$"$ denote transition metals, and X represents C or N.}
  \label{fig:mxene2}
\end{figure}

Experimentally, MXenes of Ti$_2$C~\cite{M.Naguib2012}, 
V$_2$C~\cite{M.Naguib2013}, 
Nb$_2$C~\cite{M.Naguib2013}, 
Mo$_2$C~\cite{R.Meshkini2015}, 
Ti$_3$C$_2$~\cite{M.Naguib2011}, 
Zr$_3$C$_2$~\cite{J.Zhou2016}, 
Nb$_4$C$_3$~\cite{J.Yang2016_2}, 
Ta$_4$C$_3$~\cite{M.Naguib2012}, and 
Ti$_4$N$_3$~\cite{P.Urbankowski2016} 
as well as 
TiNbC~\cite{M.Naguib2012}, 
(Ti$_{0.5}$Nb$_{0.5}$)$_2$C~\cite{M.Naguib2012},
(V$_{0.5}$Cr$_{0.5}$)$_3$C$_2$~\cite{M.Naguib2012}, 
Ti$_3$CN~\cite{M.Naguib2012}, 
Mo$_2$TiC$_2$~\cite{B.Anasori2015},
Mo$_2$ScC$_2$~\cite{R.Meshkian2017},
Cr$_2$TiC$_2$~\cite{B.Anasori2015},
Mo$_2$Ti$_2$C$_3$~\cite{B.Anasori2015},
(Nb$_{0.8}$Ti$_{0.2}$)$_4$C$_3$~\cite{J.Yang2016_2}, and  
(Nb$_{0.8}$Zr$_{0.2}$)$_4$C$_3$~\cite{J.Yang2016_2} 
have already been synthesized. 
There are comprehensive reviews on synthesis and current experimental status of 
MXenes~\cite{M.Naguib2014,M.Naguib2015,J.C.Lei2015,V.Ng2016,B.Anasori2017_1}. The 
crystal structures of 2D M$_2$X, M$_3$X$_2$, M$_4$X$_3$, 
M$'_2$M$''$X$_2$, and M$'_2$M$''_2$X$_3$ are shown in Figs.~\ref{fig:mxene1} and~\ref{fig:mxene2} .

\subsection{Structural and mechanical properties}\label{sec:structure}

The model structures for 2D MXene systems can be constructed by removing ``A'' element from the bulk 
MAX phases. Since the MAX phases have hexagonal symmetry, 
the derived MXene systems form hexagonal lattices with the same symmetry. 
As schematically shown in Fig.~\ref{fig:mxene1}(a), 2D M$_2$X MXene consists of 
trilayer sheets with a hexagonal-like unit cell, where the ``X'' layer is sandwiched by two ``M'' transition metal 
layers. Since the coordination number of a transition metal ion is often six, it is natural to assume that the transition metals in MXenes
make six chemical bonds with the neighboring X atoms and the attached chemical groups on the surfaces, 
resulting in the formation of 
M$_2$XF$_2$, M$_2$X(OH)$_2$, and M$_2$XO$_2$~\cite{M.Khazaei2013}.  

On the surfaces of M$_2$X, two types of hollow sites exist: hollow sites (A) under which there is no 
``X''   atom between the transition metal layers 
and hollow sites (B) under which there is a ``X'' atom, as indicated in the lower panel of 
Fig.~\ref{fig:mxene1}(a). 
Therefore, depending on the relative positions of the termination groups attached to 
the transition metal atoms, four different configurations are possible for the chemical termination of the 
M$_2$X system. Model 1: two functional groups locate on top of 
the two transition metals. Model 2: two functional groups are on top of hollow sites A. 
Model 3: one of the functional groups locates on top of a hollow 
site A and the other functional group locates on top of a hollow site B. 
Model 4: the two functional groups locate on top of hollow sites B. 
Figures~\ref{fig:mxene1}(b) and \ref{fig:mxene1}(c) show the schematic top and side views of models 2 and 4, 
respectively.

These four models should be considered for each type of functional groups with fully relaxing atomic positions 
to determine the most stable structure. 
As a general trend in functionalized MXenes, model 1 is energetically less stable than other three models 
and in many cases it is transformed to one of models 2--4 during structure optimization. It is also found that the relative structural stabilities 
of models 2--4 depend mainly on possible ionic state(s) of the transition metals on the 
surfaces~\cite{M.Khazaei2013}. 
If the transition metals can provide sufficient electrons to both X and the attached functional 
groups, model 2 becomes the most stable configuration~\cite{M.Khazaei2013}. 
Otherwise, either model 3 or model 4 is a more stable configuration~\cite{M.Khazaei2013}.

In order to understand the stability of functionalized MXenes, the formation energy has to be evaluated. 
First-principles calculations find that MXenes can gain a very large amount of negative 
formation energy when the surfaces are functionalized~\cite{M.Khazaei2013}. This indicates 
the formation of strong bonds between the transition metals and the attached groups, 
which encourages synthesizing MXenes with particular surface 
functionalization~\cite{M.Khazaei2013}. 
It has also been examined whether 
the surfaces of fully functionalized MXene are thermodynamically more favorable than those of 
less functionalized MXene~\cite{M.Khazaei2013,M.Ashton_2}. 
It is indeed shown theoretically that the surfaces are fully functionalized at particular chemical 
potentials in some of the MXene systems~\cite{M.Khazaei2013,M.Ashton_2}. 
Phonon frequency analyses have proved that 
many of these MXenes with full surface functionalization are locally 
stable~\cite{M.Khazaei2013,M.Khazaei2015, M.Khazaei2016_2,U.Yorulmaz2016,T.Hu2015}. 
Therefore, various fully functionalized MXenes can be realized experimentally 
in appropriate conditions although there are still some challenges.

The effect or surface functionalization on elastic constant $C_{11}$ of MXenes  
have been studied using first-principles calculations~\cite{M.Kurtoglu2012,X.H.Zha2015}. 
It is found that oxygen functionalized MXenes exhibit smaller lattice parameters and larger $C_{11}$ as 
compared to those functionalized with F or OH~\cite{X.H.Zha2015}. 
This is due to the stronger bonding between O and surface transition metals.

\begin{figure*}[t]
\centering
  \includegraphics[scale=0.32]{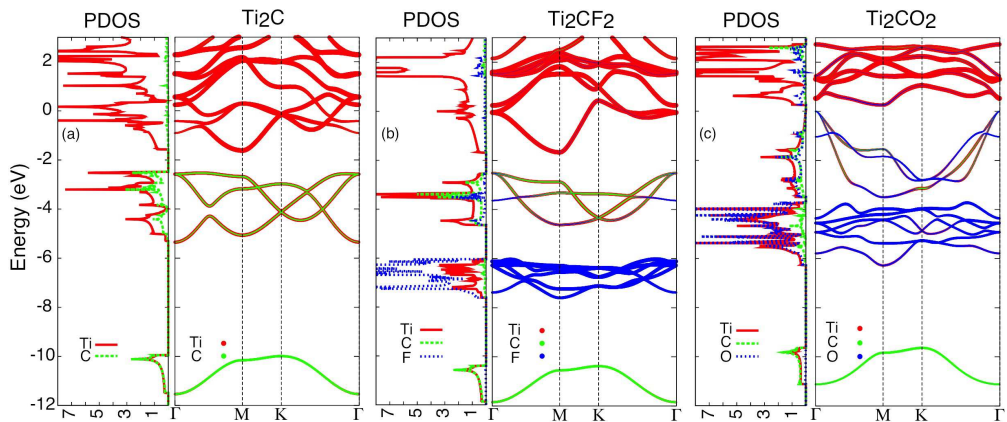}
  \caption{Projected density of states [PDOS (states/eV/cell)] and projected band structures for 
  (a) Ti$_2$C, (b) Ti$_2$CF$_2$, and (c) Ti$_2$CO$_2$~\cite{M.Khazaei2013}. 
  The Fermi energy is located at zero energy.
  }
  \label{fig:trivial}
\end{figure*}

In another theoretical study, the thickness dependence of mechanical properties of pristine 2D 
Ti$_{n+1}$C$_n$ ($n$=1, 2, and 3) has been investigated using molecular dynamics simulations.
The Young's moduluses, extracted from the slope of strain-stress curves, are found to be 597, 502, and 534 GPa 
for Ti$_2$C, Ti$_3$C$_2$, and Ti$_4$C$_3$, respectively~\cite{V.N.Borysiuk2015}, and thus 
it is the highest for the thinest MXene. 
It is also observed that Ti$_2$C and Ti$_3$C$_2$ show similar behavior under high tensile stress: 
the bond breaking begins in the outermost regions of the layer at the point of the highest local stress and 
propagates deeper into the center of the system. 
The crack propagation in these systems is accompanied by crumpling and folding of 
the layer near the growing gap \cite{V.N.Borysiuk2015}. 
In contrast, cracks in Ti$_4$C$_3$ occur first in the central region of the layer. 
At larger stress, these cracks grow in size, leading to the complete fracture of the sample with the formation 
of several Ti$_4$C$_3$ fragments~\cite{V.N.Borysiuk2015}.

The effect of F, OH, and O functionalization on the response of Ti$_{n+1}$C$_n$ ($n$=1, 2, and 3) to the tensile 
stress has also been investigated theoretically to show that Ti$_2$CO$_2$ can sustain large 
strain under biaxial and uniaxial tension much better than graphene~\cite{Z.Guo2015_1}. 
On the other hand, the pristine 2D Ti$_{n+1}$C$_n$ ($n$=1, 2, and 3) is vulnerable against the tensile strain 
because the surface atomic layer collapses easily. However, the surface functionalization can slow 
down this collapse and thus improves the mechanical flexibility~\cite{Z.Guo2015_1}. 
The unique strengthening by oxygen functionalization is attributed to the significant charge transfer from 
the inner Ti-C bonds to the outer Ti-O surface ones~\cite{Z.H.Fu2016}.

Although the F, O, and OH terminations weaken the interlayer coupling 
as compared with the pristine counterparts, the interlayer coupling is still significantly 
stronger than van der Waals bonding as specified by the fact that the binding energies of stacked 
Ti$_{n+1}$C$_n$T$_2$ (T = F, OH, or O) 
are 2--6 times larger than those of well-known graphite and MoS$_2$ 
with weak interlayer coupling~\cite{T.Hu2016}. 
Therefore, the successful exfoliation of stacked Ti$_{n+1}$C$_n$T$_2$ with binding energies in the range of 1-1.33 Jm$^{-2}$
into monolayers 
invariably requires further weakening of the interlayer coupling~\cite{T.Hu2016}. 
The interlayer coupling in the OH-terminated Ti$_{n+1}$C$_n$ is stronger than that in the 
F- or O-terminated one because of the 
formation of hydrogen bonds between the layers in the former~\cite{T.Hu2016}.

It should be noted that the experimentally obtained MXenes are not structurally perfect. 
The recent scanning transmission electron microscopy (STEM) and electron energy loss spectroscopy (EELS) 
experiments show that point defects are present on the surfaces that 
can affect the local surface chemistry~\cite{L.H.Karlsson2015,X.Sang2016}. 
The defects are formed when the surface transition metals are etched/removed by the HF in the etchant solution. 
Thus, changing the HF concentration in the etchant should vary 
the defect concentration~\cite{X.Sang2016}. This gives us the ability to control the surface properties for catalytic, 
energy storage, and other applications. However, the existence of a defect does not 
affect significantly the metallic properties of MXenes~\cite{X.Sang2016}.

Although MXenes are made by chemical exfoliation of MAX phases, their synthesis through mechanical 
exfoliation of MAX phases may also be possible. 
This is because in some of the MAX phases elastic constants satisfy $C_{11}$ > $C_{33}$, which 
indicates that the overall bondings along the $ab$ plane is stronger than the 
$c$ direction~\cite{M.Khazaei2014_1,M.Khazaei2014_2,Z.Guo2015_2}. Therefore, if 
$C_{33}$ is smaller than $C_{11}$, it might be more feasible to break the M$-$A$-$M bonds under 
appropriate mechanical tension without significantly damaging the M$-$X$-$M 
bonds~\cite{M.Khazaei2014_1,M.Khazaei2014_2}.

 \subsection{Electronic properties}\label{sec:electronic}
 
 Similar to the MAX phases, the pristine MXenes are all metallic. However, upon surface functionalization,
 some of the MXenes become semiconducting. As will be discussed below, theoretical investigations have 
 shown that the MXenes can be divided into two categories, topologically trivial~\cite{M.Khazaei2013} and 
 nontrivial~\cite{M.Khazaei2016_2,H.Weng2015_1,L.Li2016,C.Si2016,C.Si2016_2} metal/semimetal 
 or semiconductor, 
 depending on the strength of relativistic SOC in these systems.

 \subsubsection{Topologically trivial metals and semiconductors}\label{sec:m-sm}
 
While most of the functionalized MXenes are metallic,  
Sc$_2$CT$_2$ (T= F, OH, and O), Ti$_2$CO$_2$, Zr$_2$CO$_2$, and Hf$_2$CO$_2$ 
become semiconducting with the surface functionalization. The energy gaps are estimated to be 1.03, 0.45, and 1.8 eV for 
Sc$_2$CT$_2$  with T= F, OH, and O, respectively, 
0.24 eV for Ti$_2$CO$_2$, 0.88 eV for Zr$_2$CO$_2$, and 
1.0 eV for Hf$_2$CO$_2$ within general gradiente approximation (GGA) ~\cite{M.Khazaei2013}. 
The band structure calculations reveal that Sc$_2$C(OH)$_2$ has a direct band gap and the other 
semiconductors have indirect band gaps. Since Ti, Zr, and Hf are in the same group in the periodic table 
with the same number of electrons in the outermost partially filled atomic orbital shell 
([Ar]3d$^2$4s$^2$, [Kr]4d$^2$5s$^2$, and [Xe]4f$^{14}$5d$^2$6s$^2$), the corresponding MXene systems 
exhibit the similar metallic to semiconducting behavior upon the same type of functionalization. 
It is also predicted that both F and OH groups affect similarly the electronic structure of a pristine MXene 
system~\cite{M.Khazaei2013}. This is simply because each F or OH group is capable of receiving only 
one electron from the surfaces. By the same token, the O group differs from the F and OH groups because 
it demands two electrons from the surfaces to be stabilized~\cite{M.Khazaei2013}.

 In order to understand why some of the MXenes become semiconducting, the electronic structures of MXenes 
 with and without the surface functionalizations have been systematically 
 studied~\cite{M.Khazaei2013}. 
 The calculations show that pristine M$_2$X (X =C or N) systems are all metallic with the Fermi energy locating 
 at the $d$ bands of transition metal M. 
 In most MXenes, the $p$ bands of carbon/nitrogen X appear below the $d$ bands separated by a small band gap. 
With the F, OH, or O functionalization, new bands are generated 
 below the Fermi energy, resulting from F or O $p$ orbitals hybridized with M $d$ orbitals. 
 Sc$_2$CT$_2$ (T= F, OH, or O), Ti$_2$CO$_2$, Zr$_2$CO$_2$, and Hf$_2$CO$_2$ 
 turn to be semiconducting because the functionalization shifts the Fermi energy  to the center of the gap 
 between M $d$ bands and X $p$ bands. 
As examples, Fig.~\ref{fig:trivial} shows the projected density of states and projected band structures for Ti$_2$C, 
Ti$_2$CF$_2$, and Ti$_2$CO$_2$. It has also been shown that the band gap of Ti$_2$CO$_2$ and 
Sc$_2$CO$_2$ can be significantly varied by applying either strain~\cite{Y.Lee2014_1,X.F.Yu2015_2} 
or external electric field~\cite{Y.Lee2014_2,L.Li2016_2}.

 \subsubsection{Topologically non-trivial semimetals and semiconductors}\label{sec:ti}

\begin{figure}[t]
\centering
  \includegraphics[scale=0.3]{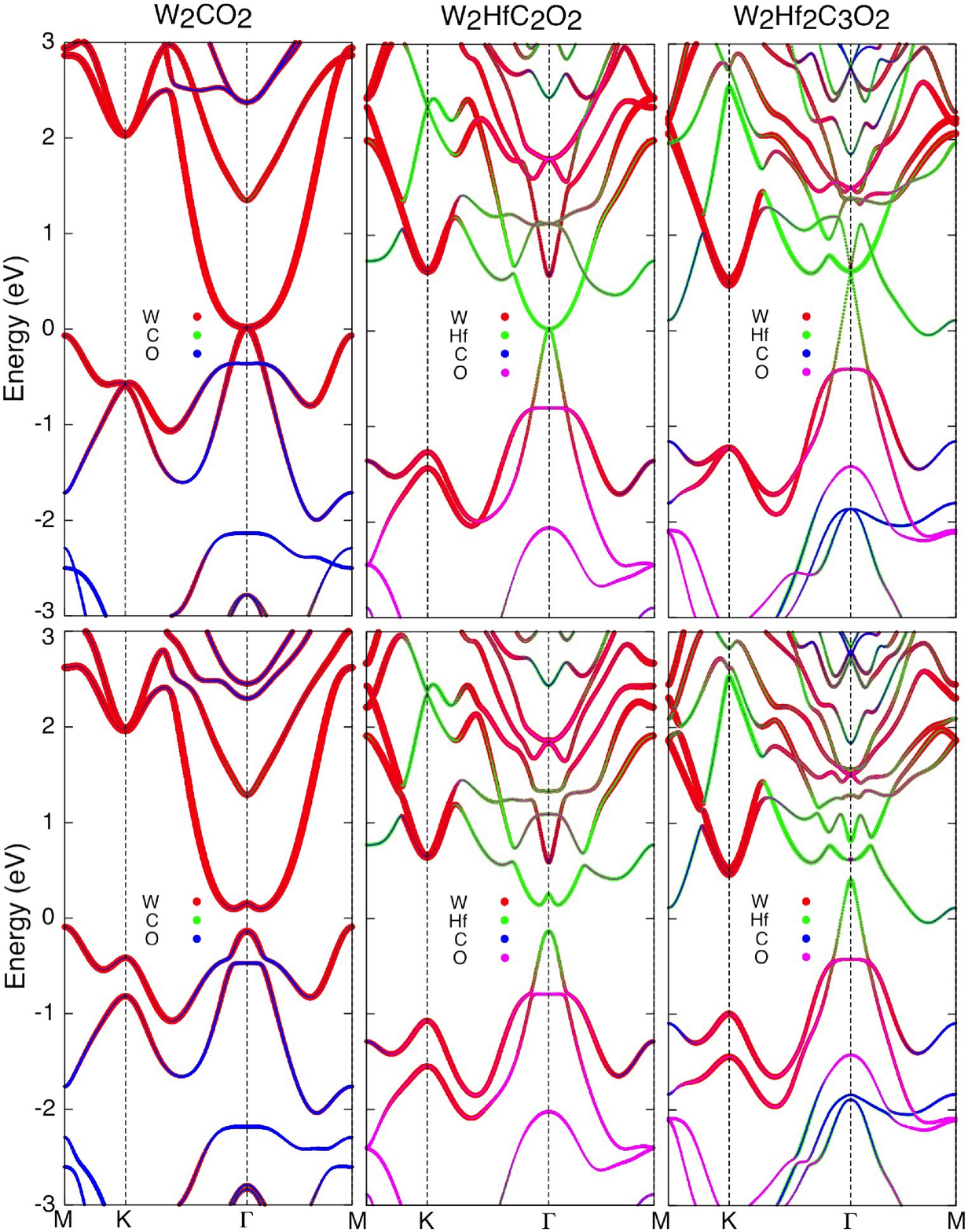}
  \caption{Top (bottom) panels: projected band structures for W$_2$CO$_2$, W$_2$HfC$_2$O$_2$, and 
  W$_2$Hf$_2$C$_3$O$_2$ calculated without (with) the relativistic spin-orbit 
  coupling~\cite{M.Khazaei2016_2,H.Weng2015_1}. 
  The Fermi energy is located at zero energy.
  }
  \label{fig:nontrivial}
\end{figure}

Since many of MXenes contain heavy 4$d$ and 5$d$ transition metals, the relativistic SOC 
affects the electronic structures significantly.   
As shown in Fig.~\ref{fig:nontrivial}, without considering the SOC, 
M$_2$CO$_2$ (M= Mo, W) and M$'_2$M$''$C$_2$O$_2$ (M$'$= Mo, W; M$''$= Ti, Zr, W) 
are semiconducting with a zero energy gap (semimetallic) because the topmost valence 
band and the lowest conduction band touch only at the $\Gamma$ point, around which the valence and conduction bands are both parabolic. The projected band structure calculations reveal that  
the bands near the Femi energy originate from $d$ orbitals of transition metals M, or M$'$ and M$''$. 
Upon considering the SOC, 
the degeneracy of the topmost valence band and 
the lowest conduction band at the Fermi energy is lifted and the band 
gap is open at the $\Gamma$ point (see lower panels in Fig.~\ref{fig:nontrivial}). 
Consequently, M$_2$CO$_2$ and M$'_2$M$''$C$_2$O$_2$ turn into semiconductors 
with indirect band gaps~\cite{M.Khazaei2016_2,H.Weng2015_1}. 
The induced band gap is larger as the SOC is larger~\cite{M.Khazaei2016_2}. 
It is  found that the band gaps can be as large as 0.194 eV (0.472 eV) for W$_2$CO$_2$ and 
0.285 (0.401 eV) for W$_2$HfC$_2$O$_2$ within the generalized
gradient approximation (hybrid functional HSE06)~\cite{M.Khazaei2016_2,H.Weng2015_1}. 
Although the similar features are observed for M$'_2$M$''_2$C$_3$O$_2$ 
(M$'$= Mo, W; M$''$= Ti, Zr, Hf), 
these systems are still semimetallic even with considering the SOC 
(see the rightmost panels in Fig.~\ref{fig:nontrivial})~\cite{M.Khazaei2016_2}.

The SOC lifts the degeneracy of the bands at the $\Gamma$ point by inducing a finite gap.
The nature of these bands around the Fermi energy can be characterized by 
the Z$_2$ topological invariant. 
Since all the M$_2$CO$_2$, M$'_2$M$''$C$_2$O$_2$, and M$'_2$M$''_2$C$_3$O$_2$ MXenes have 
the inversion symmetry, the Z$_2$ topological invariant can be simply calculated from the parity of the 
valence band wave functions at the time reversal invariant momentum (TRIM) points of the Brillouin 
zone~\cite{L.Fu2007_1,L.Fu2007_2}.  
For the hexagonal structure, the TRIM points are located at $\Gamma$: ${\bf k}_1=(0,0)$, 
$M_1$: ${\bf k}_2=(0, 0.5)$, $M_2$: ${\bf k}_3=(0.5, 0)$, and $M_3$: ${\bf k}_4=(0.5, 0.5)$, 
and therefore the $Z_2$ index $\nu\,(=0,1)$ is evaluated by $(-1)^\nu=\Pi_{i=1}^4\delta({\bf k}_i)$, 
where $\delta({\bf k}_i)=\Pi_{n=1}^N\eta_n({\bf k_i})$, $\eta_n({\bf k_i})\,(=\pm1)$ is the parity of the 
$n$th valence band at TRIM ${\bf k}_i$, and $N$ is the number of the occupied valence 
bands~\cite{L.Fu2007_1,L.Fu2007_2}. 
The trivial and nontrivial topological phases are identified with $\nu$ = 0 and 1, respectively. 
The parity analysis of the occupied bands at the TRIM points finds that $\nu$ = 1 for all the systems 
discussed here, and therefore M$_2$CO$_2$ (M= Mo, W) and M$'_2$M$''$C$_2$O$_2$ 
(M$'$= Mo, W; M$''$= Ti, Zr, Hf) are topological insulator and M$'_2$M$''_2$C$_3$O$_2$ are 
topological semimetal~\cite{M.Khazaei2016_2,H.Weng2015_1}. 
The 2D topological insulators display intriguing conducting edge states in which electrons with opposite 
spins propagate in the opposite directions and thus they are robust against non-magnetic impurities and 
disorders~\cite{M.Z.Hasan2010,X.L.Qi2011,B.Yan2012,H.Weng2015_2}.
It should also be emphasized that the band inversion occurs in these topologically non-trivial MXenes 
mostly among $d$ orbitals (more precisely, $d_z$ and $d_{xy}/d_{x^2-y^2}$ orbitals) of the transition 
metals~\cite{M.Khazaei2016_2,H.Weng2015_1}, not involving $s$ and $p$ orbitals as in the previously 
known or predicted topological insulators~\cite{M.Z.Hasan2010,X.L.Qi2011,B.Yan2012,H.Weng2015_2}.

\subsection{Surface state properties}\label{sec:surface}

\begin{figure}[t]
\centering
  \includegraphics[scale=0.265]{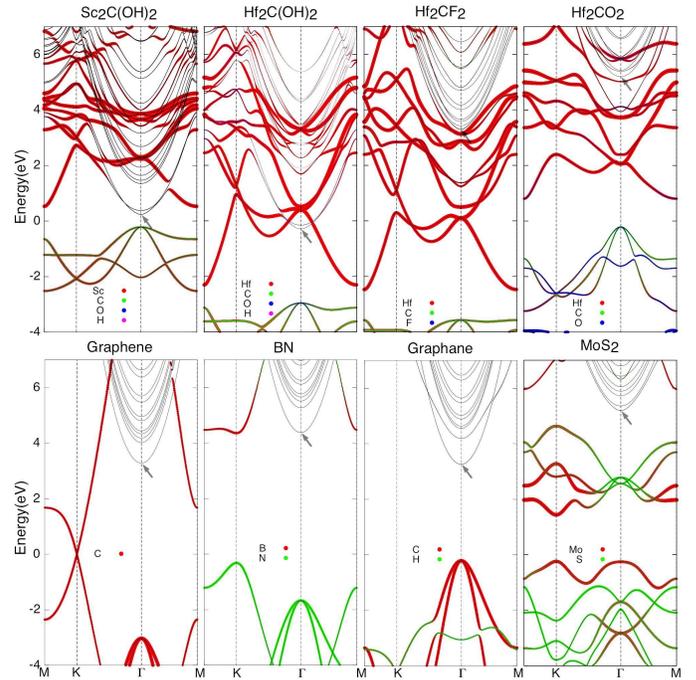}
  \caption{
Projected band structures onto each constituent atoms for eight different systems indicated 
in the figures~\cite{M.Khazaei2016_1}. 
The position of the lowest energy nearly free state (NFE) 
near the Fermi energy at the $\Gamma$ point is indicated 
by arrow. The Fermi energy is located at zero energy. 
 }
  \label{fig:nfe}
\end{figure}

The careful examination of the projected band structures of the OH-terminated MXenes 
has revealed that some of the states around the $\Gamma$ point, near and above the Fermi energy, 
do not constitute any of the composed elements~\cite{M.Khazaei2016_1}. 
As examples, the projected band structures of Sc$_2$C(OH)$_2$ and Hf$_2$C(OH)$_2$ 
are shown in Fig.~\ref{fig:nfe}. These states mainly distribute in the 
vacuum region outside the hydrogen atoms and, more interestingly, they have nearly free 
electron (NFE) characteristics with parabolic energy dispersions with respect to the crystal wave 
vector~\cite{M.Khazaei2016_1}.

\begin{figure}[t]
\centering
  \includegraphics[scale=0.35]{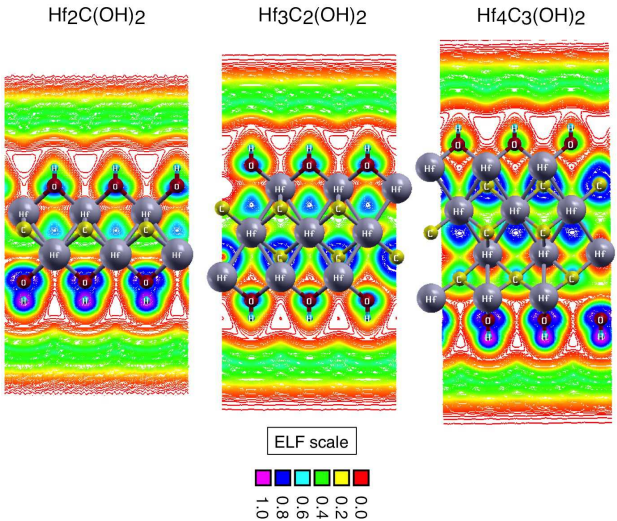}
  \caption{
Electron localization function (ELF) for Hf$_{n+1}$C$_n$(OH)$_2$ ($n$=1, 2, and 3) MXenes 
with different thicknesses~\cite{M.Khazaei2016_1}.
}
  \label{fig:elf}
\end{figure}

The electron localization function (ELF) analysis~\cite{B.Silvi1994} in Fig.~\ref{fig:elf} shows 
clearly that there exists a uniform floating electron gas above the 
hydrogen atoms in the MXenes such as Hf$_{n+1}$C$_n$(OH)$_2$ ($n=1$, 2, and 3) 
with partially occupied NFE states. 
It has been shown that 
Ti$_2$C(OH)$_2$, Zr$_2$C(OH)$_2$, Zr$_2$N(OH)$_2$, Hf$_2$N(OH)$_2$, Nb$_2$C(OH)$_2$, and 
Ta$_2$C(OH)$_2$ also possess partially occupied NFE states~\cite{M.Khazaei2016_1}. 
The lowest energy NFE states in Sc$_2$C(OH)$_2$, Ti$_2$N(OH)$_2$, and V$_2$C(OH)$_2$ are unoccupied 
and located above the Fermi energy, but in much lower energy than those of the graphene, graphane, BN, 
and MoS$_2$ (see Fig.~\ref{fig:nfe}). 
Therefore, these unoccupied NFE states in OH-terminated MXenes should be accessible more easily than those 
in other known 2D systems~\cite{M.Khazaei2016_1}.

Similarly to the OH-terminated MXenes, the F- and O-terminated MXenes also exhibit the NFE states. 
However, the NFE states in the F- and O-terminated MXenes are located at much high energies. 
For example, as shown in Fig.~\ref{fig:nfe}, 
the lowest energy NFE states of Hf$_2$CF$_2$ and Hf$_2$CO$_2$ appear around 3 and 5 eV 
above the Fermi energy, respectively. 
This already implies that the NFE states of the F- and O-terminated MXenes are not suitable for application 
purposes. 
The NFE states in the OH-terminated MXenes are found near the Fermi energy because there exist 
positively charged hydrogen atoms at the surfaces~\cite{M.Khazaei2016_1}. 
As shown in Fig.~\ref{fig:elf}, the presence of the NFE states and the location 
relative to the Fermi energy do not depend significantly on the thickness of the MXenes. 
This is simply because the thickness of the MXenes can not essentially change the nature of surface charges, 
i.e., the surface of the OH-terminated (F and O-terminated) MXenes remains positively (negatively) charged, 
independently of the thickness.

The partially occupied NFE states near the Fermi energy are sensitive to the circumstances, and 
indeed they diminish by applying large 
compressive or tensile strains, adsorbing molecules such as O$_2$, N$_2$, and 
CO, or hetrostructuring the MXenes with graphene, BN, or graphane~\cite{M.Khazaei2016_1}. 
It has also been proposed that due to their localized states that appear above
the hydrogen atoms, the OH-terminated MXenes may offer potential application toward Pb and heavy metal 
purification for environmental remediation~\cite{Q.Peng2014,J.Guo2015}.

\subsection{Optical properties}\label{sec:optical}

Very few theoretical and experimental studies exist on the optical properties of MXenes.
The experimental studies for thin films of 
Ti$_3$C$_2$T$_x$ (T: mixture of F, OH, or O) are available~\cite{J.Halim2014,M.Mariano2016}.
It has been observed that Ti$_3$C$_2$T$_x$ obtains 77$\%$ transmittance at visible light with 
wavelength of 550 nm~\cite{M.Mariano2016}.
The intercalated Ti$_3$C$_2$T$_x$ with NH$_4$HF$_2$ (Ti$_3$C$_{2.3}$O$_{1.2}$F$_{0.7}$N$_{0.2}$) 
possesses 90$\%$ transmittance, while the transmittance for thin film of the Ti$_3$AlC$_2$ MAX phase is 30$\%$, 
which is the least transparent~\cite{J.Halim2014}.  
It has also been observed that the absorbance is linearlly dependent on the thickness of the 
Ti$_3$C$_2$T$_x$ and the intercalated films~\cite{J.Halim2014}.

Theoretically, optical properties such as transmittance, absorption, and reflection can be evaluated 
from the imaginary part of the dielectric function tensor calculated as a function of the photon 
wavelength~\cite{H.Lashgari2014,Y.Bai2016}.
Thus far, the energy loss function, reflectivity, and absorption spectrum of pristine Ti$_2$C, Ti$_2$N, 
Ti$_3$C$_2$, and Ti$_3$N$_2$ have been studied. Using the reflectivity and 
energy loss curves, the plasmon energy is estimated to be 10.00, 11.63, 10.81, and 11.38 eV 
for these systems listed above, respectively, when the electric field is applied parallel to 
the surface~\cite{H.Lashgari2014}.
It has also been found that, at extremely low energy less than 1 eV,  
the reflectivity becomes 100$\%$ (less than 50$\%$) for these systems 
when the electric field is applied parallel (perpendicular) to the surface~\cite{H.Lashgari2014}. 
This indicates the capability of these systems for transmitting electromagnetic waves~\cite{H.Lashgari2014}.

In other theoretical studies, the optical properties of pristine and functionalized 
Ti$_2$C and Ti$_3$C$_2$ with F, OH, and O have been examined~\cite{Y.Bai2016}. 
It is shown 
that, in the range of infrared to ultraviolet light including visible light, the in-plane absorption coefficients are lower 
for the F and OH functionalized MXenes than the bare and O functionalized ones~\cite{Y.Bai2016}. 
By analyzing the refractivity, it is concluded that F and OH functionalized Ti$_2$C and Ti$_3$C$_2$ should exhibit 
white color~\cite{Y.Bai2016}.

\subsection{Magnetic properties}\label{sec:magnetic}

The spin-polarized density functional calculations indicate that the ground
states of the majority of the pristine and functionalized MXenes are nonmagnetic 
because of the strong covalent bonding between the transition metal and the X element as well as the 
attached chemical groups. However, the covalency of the bonds can be tuned by applying external 
strain, which results in the release of $d$ electrons and consequently 
the appearance of magnetism even in the otherwise nonmagnetic systems~\cite{S.Zhao2014}.

Many of the pristine MXenes are intrinsically magnetic. For example,
pristine Ti$_2$C and Ti$_2$N exhibit nearly half-metallic ferromagnetism with the magnetic 
moments of 1.91 and 1.00 $\mu$B per formula unit. 
The nearly half metallic ferromagnetic monolayer Ti$_2$C turns into a perfect half-metal, 
to a spin-gapless semiconductor, and then a metal, under the continuous increase of the biaxial strain, 
although the monolayer Ti$_2$N remains nearly half-metallic under the similar biaxial strain~\cite{G.Gao2016}.

The monolayers of V$_2$C and V$_2$N are antiferromagnetic and nonmagnetic metals, respectively.
The biaxial tensile and compressive strains induce large magnetic moment in 
monolayer V$_2$C and V$_2$N, respectively~\cite{G.Gao2016}. 
Upon the F or OH functionalization,V$_2$C becomes a small-gap antiferromagnetic semiconductor~\cite{J.Hu2014}.

\begin{figure}[t]
\centering
  \includegraphics[scale=0.4]{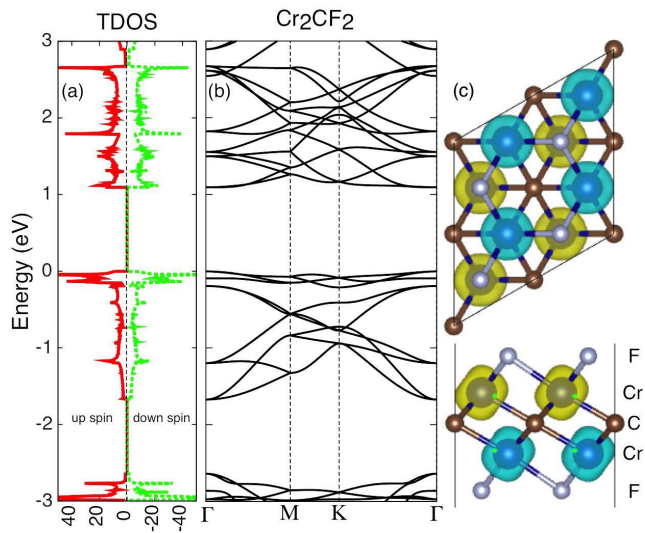}
  \caption{
  Electronic structure of antiferromagnetic Cr$_2$CF$_2$ obtained using the GGA: 
  (a) spin projected density of states 
  [TDOS (states/eV/cell)], (b) band structure, and (c) top and side views of spin density. 
  The Fermi energy is located at zero energy in (a) and (b). 
  Blue and yellow colors indicate spin up and down densities, respectively, in (c). 
  }
  \label{fig:magnetic}
\end{figure}

Pristine Cr$_2$C and Cr$_2$N are both magnetic. The magnetism results from $d$ electrons of Cr. 
The Cr$_2$C is half-metallic ferromagnetic. However, with the F, Cl, OH, or H functinalization, 
Cr$_2$C is transformed to an antiferreomagnetic semiconductor~\cite{C.Si2015}.
The electronic structure of Cr$_2$CF$_2$ is shown in Fig.~\ref{fig:magnetic}.
Cr$_2$N has a larger number of valence electrons than Cr$_2$C. Hence, 
in contrast to Cr$_2$C, the ground state of Cr$_2$N is antiferreomagnetic. 
However, the half-metallicity appears in Cr$_2$N upon various surface functionalization with F, O, or 
OH~\cite{G.Wang2016}. 
This is important because the synthesize of the pristine half-metallic MXenes such as
Cr$_2$C is difficult when the etching technique is employed. 
It is also found that the energy difference between the magnetic and non-magnetic states 
of the Cr-based MXenes is so large that these systems might probably keep their magnetism 
up to near room temperature~\cite{M.Khazaei2013}. Thus, these MXenes are promising materials for 
spintronic devices.

Recently, non-symmetrically functionalized graphene has 
been realized and named as Janus graphene~\cite{L.Zhang2013}. 
Janus materials possess distinct surfaces because their opposite faces are 
functionalized with two different chemical groups.  
Motivated by this, the magnetic properties of MXenes' version of Janus graphene has been studied~\cite{J.He2016,M.Je2016}. 
It is shown that Cr$_2$CFCl, Cr$_2$CHF, Cr$_2$CFOH, and V$_2$CFOH can be antiferromagnetic 
semiconductors~\cite{J.He2016}.

There are also attempts to induce magnetism in nonmagnetic 
MXenes such as Sc$_2$CT$_2$ (T= F, OH, and O) by doping with Ti, V, Cr, or Mn. 
It is found that Mn and Cr are promising dopants for achieving magnetic 
Sc$_2$CT$_2$~\cite{J.Yang2016}.

More recently, the magnetic properties of ordered double transition metal carbides MXenes,
Cr$_2$M$''$C$_2$T$_2$ (M$''$= Ti and V, T= F, OH, and O), have been investigated~\cite{J.Yang2016_3}.
It is found that Cr$_2$M$''$C$_2$T$_2$ can be nonmagnetic, anti-ferromagnetic, or ferromagnetic and 
either semiconducting or metallic, depending on M$''$ and T. 
It is also shown that Cr$_2$TiC$_2$O$_2$ is nonmagnetic, whereas 
Cr$_2$TiC$_2$F$_2$ and Cr$_2$TiC$_2$(OH)$_2$ are anti-ferromagnetic, and 
Cr$_2$VC$_2$(OH)$_2$, Cr$_2$VC$_2$F$_2$, and Cr$_2$VC$_2$O$_2$ are
ferromagnetic. 
The Curie temperatures of Cr$_2$VC$_2$(OH)$_2$ and Cr$_2$VC$_2$F$_2$ are estimated to 
be up to 618.36 and 695.65 K, respectively~\cite{J.Yang2016_3}. Another intensive study 
has found robust ferromagnetism in Hf$_2$MnC$_2$O$_2$ and Hf$_2$VC$_2$O$_2$, as well as 
in Ti$_2$MnC$_2$T$_2$ (T= F, OH, and O). In these systems, a large magnetic moment of 3--4 $\mu$B/cell 
and a high Curie temperature of 495--1133 K are predicted~\cite{L.Dong2017}.

 \begin{figure}[t]
\centering
  \includegraphics[scale=0.29]{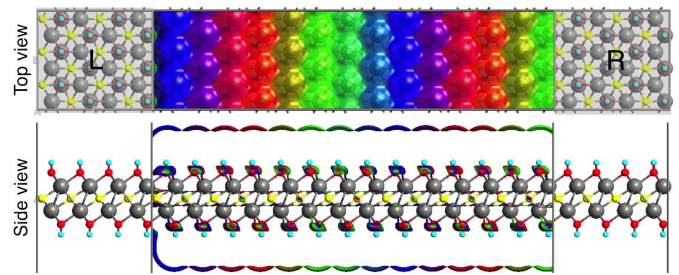}
  \caption{
  Top and side views of one of the eigenvectors of transmission matrix with 
 NFE characteristic for Hf$_2$C(OH)$_2$~\cite{M.Khazaei2016_1}. L and R indicate the left and right electrodes, 
respectively. Since the transmission eigenvector is complex, the absolute value of the eigenvector 
is shown by the isosurface and the phase is indicated by the color of the isosurface.
}
  \label{fig:transport}
\end{figure}

\subsection{Transport properties}\label{sec:transport}

There are very few theoretical studies on electron transport properties of the MXenes.
The coherent transport calculations using the non-equilibrium Green's function (NEGF)
scheme have shown that the metallic MXenes are highly conductive~\cite{M.Khazaei2016_1,G.R.Berdiyorov2015}. 
According to the analyses of pristine and functionalized Ti$_3$C$_2$,  the type of the surface 
functionlization has a considerable impact on the electron transport. 
For example, the current at a given bias voltage in
Ti$_3$C$_2$F$_2$ can be up to 4 times larger than that for pristine Ti$_3$C$_2$~\cite{G.R.Berdiyorov2015}. 
The current increase is due to the emergence of extended electronic states and 
also due to the smaller variation of electrostatic profile~\cite{G.R.Berdiyorov2015}. 
Through first-principles calculations, it has been demonstrated that the conduction in 
stacked of Ti$_3$C$_2$T$_2$ (T= F, OH, or OH) is significantly anisotropic due to different effective masses of electrons and 
holes along different lattice directions~\cite{T.Hu2015_2}. such anisotropic electronic conduction was evidenced by in situ I-V measurements, in which the in-plane electrical conduction is a least one order of magnitude higher than that vertical to the basal plane~\cite{T.Hu2015_2}.
It is also reported that the half metallicity of Cr$_2$NO$_2$ is robust with 100$\%$ 
polarization for a wide range of applied voltages~\cite{G.Wang2016}.

Electron transport properties of the MXenes with the partially occupied NFE states have also been 
studied~\cite{M.Khazaei2016_1}. 
The calculations have shown the signature of the NFE states in electron conduction~\cite{M.Khazaei2016_1}. 
In general, the better electron transport is expected for the MXenes with the NFE states 
because these states extend above the MXene surfaces (see Fig.~\ref{fig:transport}) 
and can conduct electrons through the nearly free channels without being significantly scattered 
by surface vibrations at finite temperatures. The transmission analysis indeed clearly shows
that the NFE states can act as both hole and electron channels at low bias voltages, thus
suitable for applications in low-power nanoelectronic devices~\cite{M.Khazaei2016_1}.

The effect of NH$_3$, H$_2$, CH$_4$, CO, CO$_2$, N$_2$, NO$_2$, and O$_2$ absorption on the electron
transport of Ti$_2$CO$_2$ has also been investigated theoretically to exploit its potential 
applications as gas sensor or capturer~\cite{X.F.Yu2015_1}. Among all these gas molecules, only NH$_3$ 
could be chemisorbed on Ti$_2$CO$_2$ with apparent charge transfer of 0.174~e. The electron transport 
exhibits distinct responses with a drastic change of I-V characteristic when NH$_3$ is adsorbed on 
Ti$_2$CO$_2$. Thus, Ti$_2$CO$_2$ could be a promising candidate for NH$_3$
sensor with high selectivity and sensitivity~\cite{X.F.Yu2015_1}.

\subsection{Applications}\label{sec:appl}

There have been many proposed applications for the 2D MXenes in literature. 
Here, we shall focus only on those with electronic and energy harvesting applications.

\subsubsection{Low work function electron emitters}\label{sec:wf}

Various metallic substrates of transition metal carbides and nitrides provide the high
stability, high melting point, and relatively low work function that have 
increasingly attracted attentions as suitable field emitters. Considering the properties of 
the MXenes, members of this family can be good candidates as materials 
with low work function because of the following two reasons. First, the MXenes offer the 
tunability of the work function by choosing the proper transition metal as well as the X
element, usually carbon or nitrogen. In addition, such compositional tunability provides a
possibility to adjust other properties such as thermal, mechanical, and chemical stabilities, 
and toxicity. Second, during the synthesis of MXenes, their surfaces are 
functionalized naturally, which changes the electrostatic potential near the surfaces and 
affects the electronic structures, especially, giving rise to the shift of the Fermi level. These 
two features have significant effects on the work function~\cite{M.Khazaei2015}.

The work function is defined as the energy difference between the Fermi energy and
the vacuum level (electrostatic potential away from the surface). From the work function
calculations of M$_2$C and M$_{10}$C$_9$ (M= Sc, Ti, Zr, Hf, V, Nb, Ta) as well as M$'_2$N and 
M$'_{10}$N$_9$ (M$'$= Ti, Zr, and Hf) functionalized with F, OH, and O, it is found that the work 
functions of the bare MXenes and the functionalized ones with F, OH and O are distributed in 
the range of 3.3$-$4.8, 3.1$-$5.8, 1.6$-$2.8, and 3.3$-$6.7 eV, respectively~\cite{M.Khazaei2015}. 
It is also shown that work functions always decrease by the functionalization with OH. 
This is in remarkable contrast with the systems functionalizated with F or O, in which the work functions 
increase or decrease depending on the type of transition metal. 
Interestingly, 
M$_2$C(OH)$_2$, M$_{10}$C$_9$(OH)$_2$, M$'_2$N(OH)$_2$ and M$'_{10}$N$_9$(OH)$_2$ 
all exhibit ultralow work functions, as compared with the corresponding bare and the functionalized ones 
with F and O~\cite{M.Khazaei2015}. 
The thickness dependences of the work function for Ti$_{n+1}$C$_n$ (n=1--9) functionalized with 
F, OH and O has been systematically studied, and shown that, regardless of the thickness, 
a family of Ti$_{n+1}$C$_n$(OH)$_2$ exhibits ultralow work function~\cite{M.Khazaei2015}.

\begin{figure}[t]
\centering
  \includegraphics[scale=0.35]{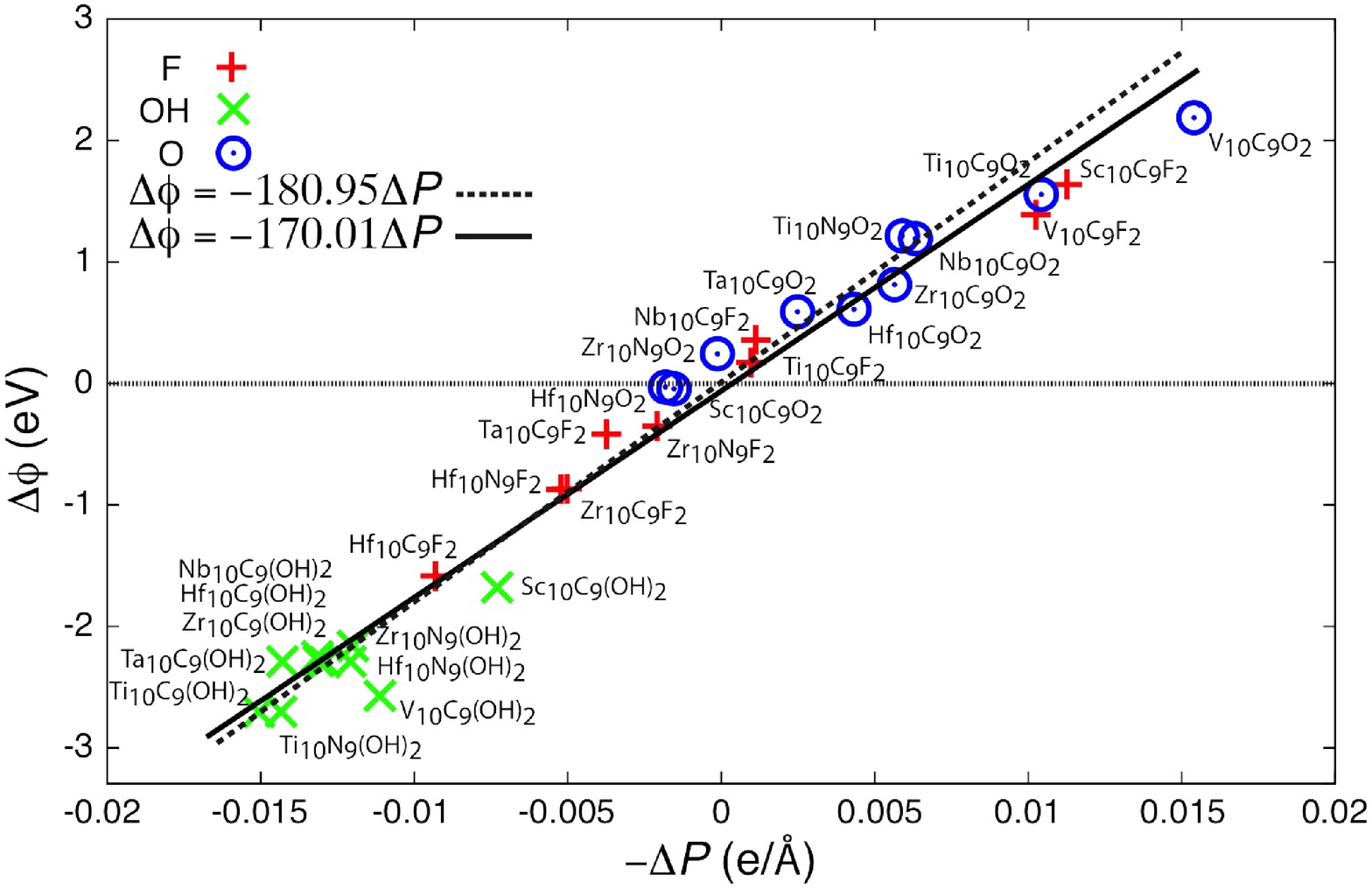}
  \caption{The change in the work function ($\Delta\Phi$) of 
  M$_{10}$C$_9$ and M$'_{10}$N$_9$ (M= Sc, Ti, Zr, Hf, V, Nb, Ta; M$'$= Ti, Zr, Hf) 
  functionalized with F, OH, and O as a function of the change in the surface dipole 
  moment ($\Delta$P)~\cite{M.Khazaei2015}. 
  The solid line 
  ($\Delta\Phi = -170.01\Delta P$) indicates the linear fit to the results. 
  The dashed line ($\Delta\Phi = -180.95\Delta P$) is the linear relation 
  reported previously for three-dimensional bulk systems~\cite{T.C.Leung2003}.}
  \label{fig:workfunction}
\end{figure}

The work function properties of MXenes can be well understood by the surface
dipole moments because the change of the work function  ($\Delta\Phi$) upon
functionalization is linearly correlated with the change of the surface dipole moment ($\Delta P$). 
It has been shown in Ref.~\cite{T.C.Leung2003} that 
$\Delta\Phi = -e/\epsilon_0\Delta P = -180.95 \Delta P$ with $\Delta\Phi = \Phi-\Phi_0$ and 
$\Delta P=p-p_0$. Here, $e$ (C) is the charge of electron, 
$\epsilon_0$ (CV$^{-1}$ \AA$^{-1})$ is the vacuum permittivity, 
and $\Phi_0(\Phi)$ and $p_0$ ($p$) are the work function and the surface dipole moment without (with) 
functionalization, respectively. 
 The positive (negative) value of $\Delta\Phi$ corresponds to the increase (decrease) of the work function 
 when the surface is functionalized. 
 As shown in Fig.~\ref{fig:workfunction},
 MXenes also follow this rule because it is found that $\Delta\Phi$ correlates with $\Delta P$ almost linearly 
 with the slope of $-170.012$ V\AA~\cite{M.Khazaei2015}.

 In general, three main factors control the surface dipole moment upon deposition of any element: 
 i) the redistribution of electron charge between the surface and the adsorbates, 
 ii) the surface relaxation caused by the adsorbates, and iii) the polarity of the 
 adsorbates~\cite{M.Khazaei2015,T.C.Leung2003}. Further analysis on the MXenes indicates that 
the work function of the F and O functionalized MXenes is controlled by two factors, i.e., the
dipole moment induced by the charge transfer between F/O and the substrate, and the
change of the total surface dipole moment caused by the surface relaxation upon 
functionalization. Besides these two factors, the intrinsic dipole moment of the OH group plays the most important 
role in determining the total dipole moment, which consequently 
justifies the ultralow work function found in the OH terminated MXenes~\cite{M.Khazaei2015}.

\subsubsection{Promising catalysts and photocatalysts for hydrogen evolution}

Hydrogen is one of the most promising alternative sources of clean energy 
because the burning of hydrogen produces no pollution~\cite{renewable}. 
Despite its abundance in water and many hydrocarbons, hydrogen is not found in gas form 
with high density in nature.
A hydrogen evolution reaction (HER) is the production of hydrogen through the 
electrocatylysis/photocatylysis process of water or any other inorganic molecules containing hydrogen.

The platinum is one of the most efficient catalysts for hydrogen evolution. However, the usage of platinum is restricted because of its high cost and resource limitation. 
Recently, MXenes have attracted much attention both experimentally and theoretically because MXenes can 
be efficient catalysts or photocatalysts for HER. 
HER activities of Ti$_2$CT$_x$, Ti$_3$C$_2$T$_x$, and Mo$_2$CT$_x$ (T: mixture of O and OH) have been measured experimentally~\cite{G.Fan2017,Z.W.Seh2016,K.D.Fredrickson2016}.
It is found that Mo$_2$CT$_x$ exhibits a higher HER activity than Ti$_2$CT$_x$ for hydrogen production 
from water~\cite{Z.W.Seh2016}, while Ti$_3$C$_2$T$_x$ can enhance the 
catalytic activity for hydrogen production from ammonia borane~\cite{G.Fan2017}. 
Theoretically, it is demonstrated that many MXenes are conductive at the standard condition (pH=0).
Therefore, this enables MXenes to transfer charge during the HER. 
Moreover, it is shown theoretically that the Gibbs free energy for the hydrogen absorption on 
some of the O-terminated MXenes such as Ti$_2$CO$_2$ and W$_2$CO$_2$ 
is close to the ideal value of 0 eV~\cite{C.Ling2016_2,G.Gao2016_2}. 
Considering the Gibbs free energy and the 
number of electrons captured by oxygen atoms on the O-terminated MXenes, 
it is predicted that O-terminated MXenes 
with 0.895--0.977 electron captured per an oxygen atom 
can be promising 2D catalysts for hydrogen production~\cite{C.Ling2016_2}.   


More recent theoretical studies have shown that M$_2$CO$_2$ (M= Ti, Zr, and Hf) exhibits 
the band gap in the range of 0.92--1.75 eV (calculated using the HSE06 method)~\cite{H.Zhang2016} 
with very good light absorbance in the wavelength from 300 to 500 nm~\cite{Z.Guo2016}. 
Moreover, it is found that Zr$_2$CO$_2$ and Hf$_2$CO$_2$ exhibit high and directionally anisotropic 
carrier mobility~\cite{Z.Guo2016}. 
The large and anisotropic carrier (electron and hole) mobility in these 
systems facilitate the migration and separation of photogenerated electron-hole pairs~\cite{Z.Guo2016}. 
Therefore, Zr$_2$CO$_2$ and Hf$_2$CO$_2$ can be used as 
photocatalyst for hydrogen generation from water~\cite{Z.Guo2016}.

\subsubsection{Energy conversion: thermoelectric devices}\label{sec:thermo}

Owing to their intrinsic ceramic nature, MXenes may be suitable as thermoelectric materials for 
energy conversion applications at high temperatures. The performance of  thermoelectric material 
is quantified through the dimensionless figure of merit ZT given as $S^2\sigma T/K$, where $S$, $\sigma$, 
$T$, and $K\,(=k_{\rm l}+k_{\rm e})$ are the Seebeck coefficient, electrical conductivity, temperature, and 
thermal conductivity with both lattice ($k_{\rm l}$) and electronic ($k_{\rm e}$) contributions, respectively. 
Generally, ZT can be maximized 
when the power factor ($S^2\sigma$) is maximized and at the same time the thermal conductivity is minimized. 
The thermal conductivity can be minimized efficiently by enhancing the phonon scattering in the presence of 
edges, interfaces, grain boundaries, and embedded nanostructures~\cite{M.Zebarjadi2012}. 
However, maximizing $S^2\sigma$ is not straightforward because both $S$ and $\sigma$ are strongly coupled to 
the electronic structure of the system and usually behave inversely: materials that have high Seebeck 
coefficient have poor electrical conductivity, and vice versa. Therefore, the balance between 
the Seebeck coefficient and the electrical conductivity 
at a particular p- or n-type carrier concentration is required so as to maximize the power factor.

\begin{figure}[t]
\centering
  \includegraphics[scale=0.305]{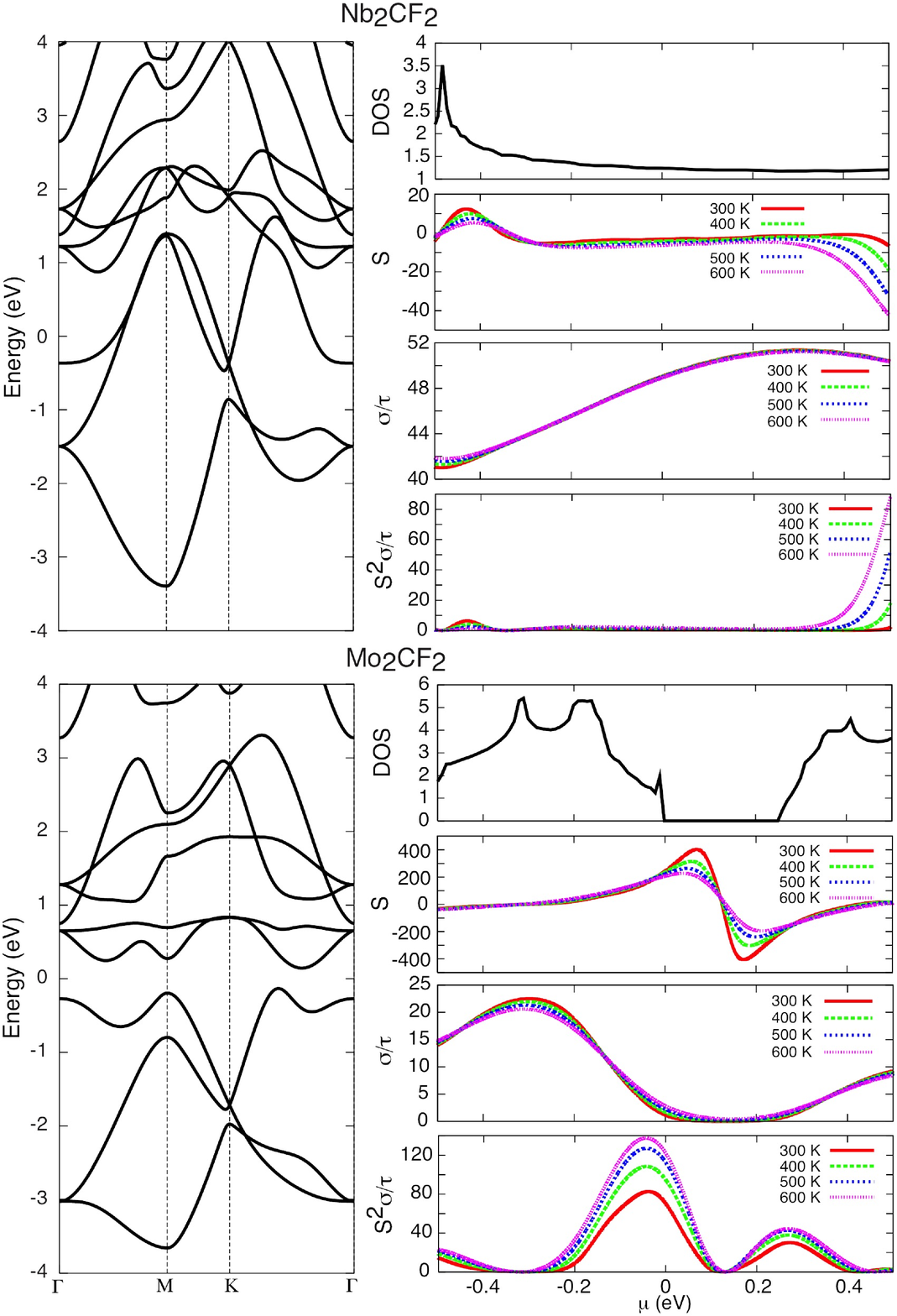}
  \caption{Band structure, density of states (DOS [states/eV/cell]), Seebeck coefficient 
  (S [$\mu$VK$^{-1}$]), electrical conductivity 
  ($\sigma/\tau$ [10$^{17}\Omega^{-1}$cm$^{-1}$s$^{-1}$], 
 and power factor (S$^2\sigma/\tau$ [10$^{14}\mu$Wcm$^{-1}$K$^{-2}$s$^{-1}$]) 
 as a function of chemical potential $\mu$ at 
 various temperatures (indicated in the figures) for Nb$_2$CF$_2$ (upper panels) and 
 Mo$_2$CF$_2$ (lower panels)~\cite{M.Khazaei2014}. 
 Here, $\tau$ is the relaxation time. 
 The band structure 
 and DOS are calculated at zero temperature. 
 $\mu$=0 corresponds to the Fermi energy at zero temperature in DOS, Seebeck coefficient, 
 electrical conductivity, and power factor.
  }
 \label{fig:thermo}
\end{figure}

 Based on the Boltzmann theory and first-principles electronic structure calculations, 
 the power factor $S^2\sigma$ of monolayer and multilayer M$_2$C (M = Sc, Ti, V, Zr, Nb, 
 Mo, Hf, Ta) and M$_2$N (M= Ti, Zr, Hf) MXenes functionalized with F, OH, and O groups have already 
 been explored~\cite{M.Khazaei2014}. It has been shown that the metallic (semiconducting) MXenes are poor 
 (good) thermoelectric materials~\cite{M.Khazaei2014}. 
 For example, as shown in Fig.~\ref{fig:thermo} for Nb$_2$CF$_2$, the metallic MXenes exhibit 
 excellent electrical conductivity, but very poor Seebeck coefficient, 
 consequently having very poor power factor. 
 On the other hand, the semiconducting MXenes show relatively poor electric conductivity, 
 but excellent Seebeck coefficient, resulting in relatively good power factor. 
 Among the semiconducting MXenes, Mo$_2$CF$_2$ (see Fig.~\ref{fig:thermo}) acquires superior power factor 
 over other MXenes and thus it is potential thermoelectric material in the MXene family~\cite{M.Khazaei2014}. 
 The exceptionally superior thermoelectric property of Mo$_2$CF$_2$ is attributed to the 
 peculiar band structure near the band edge, where flat and dispersive bands are merged. This type of  
 band shapes allows for a large Seebeck coefficient and simultaneously a good electrical conductivity 
 at low carrier concentrations~\cite{A.F.May2009,D.J.Singh1997,D.J.Singh2010}.

 The thermal conductivity of various MXenes, e.g., M$_2$CO$_2$ 
 (M= Ti, Zr, Hf)~\cite{A.N.Gand2016,X.H.Zha2016_2}, 
 Sc$_2$CT$_2$ (T= F, O, OH)~\cite{S.Kumar2016,X.H.Zha2016_1}, and Mo$_2$C~\cite{X.H.Zha2016_3}, 
 have also been studied by taking into account both phonon and electron contributions.  
 These MXenes are all semiconducting except for pristine Mo$_2$C that is metallic. 
  The thermal conductivity is in general anisotropic along the armchair and zigzag directions~\cite{X.H.Zha2016_2}. 
  Among M$_2$CO$_2$ (M= Ti, Zr, Hf), Ti$_2$CO$_2$ and Hf$_2$CO$_2$ exhibit the lowest and highest 
  lattice thermal conductivities, respectively, in the temperature range of 300--700 K. 
  Consequently, Ti$_2$CO$_2$ achieves a higher figure of merit ZT for thermoelectric application 
  (ZT= 0.45 for $14.0\times10^{20}$ cm$^{-3}$ $n$-dopeing and ZT= 0.27 for $1.7\times10^{20}$ 
  cm$^{-3}$ $p$-doping~\cite{A.N.Gand2016}) 
 because of the low thermal conductivity and the high carrier mobility~\cite{A.N.Gand2016,X.H.Zha2016_2}. 
 The room temperature thermal conductivity of Hf$_2$CO$_2$ is 62.12--131.2 (27.63--53.03) Wm$^{-1}$K$^{-1}$ 
 along the armchair (zigzag) flake length of 1--100$\mu$m~\cite{X.H.Zha2016_2}. 
 Among Sc$_2$CT$_2$ (T= F, O, OH), Sc$_2$C(OH)$_2$ is predicted to be a candidate for 
  thermoelectric application at elevated temperatures more than 300 K due to a relatively good 
 Seebeck coefficient, high electrical conductivity, and very low lattice thermal conductivity~\cite{S.Kumar2016}. 
 At room temperature, the thermal conductivity of Sc$_2$C(OH)$_2$ 
 along the armchair (zigzag) direction can be as large as 123--245 (78.5--148) Wm$^{-1}$K$^{-1}$ 
 when the flake length increases from 1 to 50 $\mu$m \cite{X.H.Zha2016_1}. 
 The room temperature thermal conductivity of 
 Mo$_2$C is also evaluated to be 48.4 (64.7) Wm$^{-1}$K$^{-1}$ for the armchair (zigzag) flake length of 
 100 $\mu$m~\cite{X.H.Zha2016_3}.

\subsubsection{Energy storage: hydrogen storage, ion batteries, and supercapacitors }\label{sec:hs}

There is a worldwide contest to develop alternative clean energy resources. 
In this regard, alkali metal batteries, supercapacitors, fuelcells, and hydrogen fuels are among most 
important candidates, although each of these faces its challenges. 
For instance, there is a major challenge to find appropriate materials with high storage capacity. 
2D materials with large surface areas are most promising for hydrogen 
storage~\cite{M.Khazaei2009,N.S.Venkataramanan2009}. 
The calculations show that pristine Sc$_2$C, Ti$_2$C, and V$_2$C might be used as 
hydrogen storage media~\cite{Q.Hu2013,Q.Hu2014}. 
Sc$_2$C with the largest surface area among all MXenes possesses 
the 9.0 wt.$\%$, 
which is higher than gravimetric storage capacity target, 5.5 wt.$\%$ by 2015, 
set by United States Department of Energy.
It has also been shown that the hydrogens are bound to Sc$_2$C  by 
chemical, physical, and Kubas-type interactions with the binding energy of 4.703, 0.087, and 0.164 eV, 
respectively~\cite{Q.Hu2014}. Note, however, that when the hydrogens are bound with the surfaces 
strongly, it is very difficult to take them out from the storage media. Therefore, these hydrogens 
cannot be used for the energy production.

Li-ion batteries can produce high gravimetric energy density and high voltage as large as 
110--160 Wh/Kg~\cite{D.Linden2001} and 3.6 V~\cite{J.M.Tarascon2010}, respectively. 
Therefore, there are extensive theoretical and experimental studies to design novel 
alkali-based batteries with high gravimetric energy density and 
voltage~\cite{M.S.Islam2014,K.M.Bui2016,K.M.Bui2015,K.M.Bui2012}. 
Recently, the intercalation between 2D Ti$_3$C$_2$T$_x$ MXene layers and several ions 
such as Li$^+$, Na$^+$, Mg$^{2+}$, K$^+$, NH$_4^+$, and Al$^{3+}$ has been 
reported to induce high capacitance up to 350 F/cm$^3$~\cite{M.R.Lukatskaya2013}. 
Moreover, Li ion batteries in Ti$_2$C, Ti$_3$C$_2$, Ti$_3$CN, and TiNbC have been experimentally 
examined~\cite{J.Come2012,O.Mashtalir2013}. It is found that Ti$_3$C$_2$ shows excellent Li-ion 
steady-state capacity of 410 mAhg$^{-1}$ at a one-cycling rate~\cite{O.Mashtalir2013}.

These important experimental achievements have triggered many further theoretical studies 
on supercapacity and alkali-metal ion battery properties of MXenes~\cite{Y.Xie2014_1,J.Hu2014,Q.Tang2012,M.Ashton2016,X.Ji2016,Y.Ando2016,D.Er2014,Y.X.Yu2016,E.Yang2015,J.Hu2016,Y.Xie2016_2,C.Eames2014,L.Bai2016,F.Li2016,H.Pan2015,J.Zhu2016,X.Chen2016}. 
In these theoretical studies, the effect of functionalization on mobility of various alkali-metals and ions 
(Li, Na, K, Ca) in pristine MXenes, Ti$_2$C, Zr$_2$C, V$_2$C, Nb$_2$C, Cr$_2$C, Ti$_2$N, V$_2$N, 
Ta$_2$N, Ti$_3$C$_2$, V$_3$C$_2$, and V$_3$N$_2$, as well as functionalized MXenes with F, O, OH, O, S, 
Se, and Te have been systematically studied.  
For instance, M$_2$C (M= Sc, Ti, V, and Cr) has gravimetric capacity over 400 mAh/g, which is higher than 
the gravimetric capacity of graphite~\cite{M.Ashton2016}. 
The simulations on Li storage in Ti$_3$C$_2$ indicate that the pristine MXenes have low diffusion 
barrier (0.07 eV) and high Li storage capacity \cite{Q.Tang2012}. 
It should be recalled that the Li storage capacity of graphite is 372 mAh/g in LiC$_6$~\cite{Q.Tang2012} and 
the diffusion barrier for Li in graphite is $\sim$0.3 eV~\cite{K.Persson2010_1,K. Toyoura2010,K.Persson2010_2}. 
These facts suggest that the MXenes may be superior to graphite electrodes. 
However, in functionalized Ti$_3$C$_2$, due to steric hindrance induced by F, OH, and O groups, 
Li ions face higher diffusion barriers as large as 0.36~\cite{Q.Tang2012}, 1.02~\cite{Q.Tang2012}, and 
0.62~\cite{M.Ashton2016} eV, respectively. 
Consequently, the Li storage in the functionalized MXenes with F, O, and OH decreases 
as compared with the pristine MXenes. 
Various theoretical studies have shown that MXenes, particularly pristine ones, are overall promising in increasing 
the battery performance~\cite{Y.Xie2014_1,Q.Tang2012}.

\begin{figure}[t]
\centering
  \includegraphics[scale=0.43]{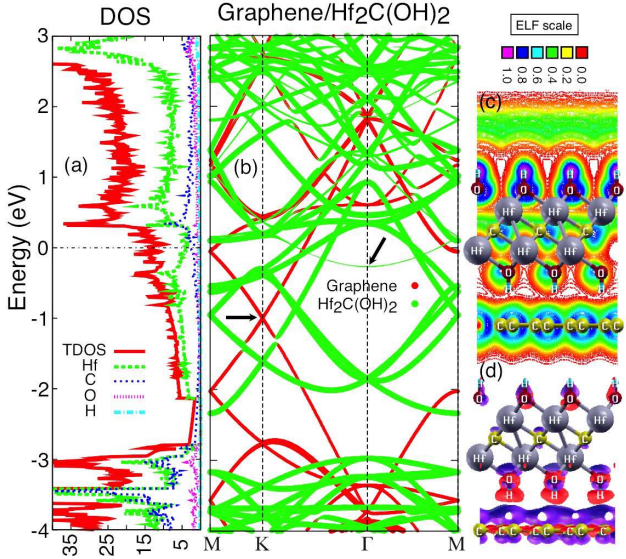}
  \caption{
  (a) Density of states (DOS) in states/eV/cell and (b) projected band structure of
graphene/Hf$_2$C(OH)$_2$ heterostructure~\cite{M.Khazaei2016_1}. 
Arrows in (b) indicate the position of the Dirac cone at the 
K point and the nearly free electron band band at the $\Gamma$ point. The Fermi energy is located at zero 
energy. (c) Electron localization function (ELF) counter plot and (d) charge transfer ($\Delta\rho$) isosurfaces 
with $\pm$0.005 e/\AA$^3$ for graphene/Hf$_2$C(OH)$_2$~\cite{M.Khazaei2016_1}. 
The excess and depleted charges (electrons) are shown in dark orchid and deep pink colors, respectively, in (d).   
  }
 \label{fig:hetero}
\end{figure}

\section{Nanoribbon, nanotube, and heterostructure MXenes}\label{sec:nano}

Although there is no experimental report on the formation of nanotubes, nanoribbons, 
or hetreostructures of MXenes, there exist many positive signs to expect that they might be realized
experimentally. In fact, experiments have observed that MXenes can be 
formed in the conical scrolls~\cite{M.Naguib2011}, and demonstrated that MXenes can be used as 
filler in polymer composites~\cite{X.Zhang2013}. 
Moreover, the chemical vapor deposition technique can be employed to synthesize MXenes~\cite{C.Xu2015}. 
On the other hand, there have been several theoretical studies on nanoribbon, nanotube, and heterostructure 
MXenes. For example, the electronic structure and formation energy calculations on pristine and 
functionalized Sc$_2$C nanotubes indicate that the stability and the band gap of the MXene nanotubes 
can be controlled significantly by their radii~\cite{X.Zhang2015_1}. 
Another study has shown that the strain energies of Ti$_2$C nanotubes are always positive, 
whereas those
of Ti$_2$CO$_2$ nanotubes become negative when the diameter is beyond 2.5 nm. 
This indicates that 2D Ti$_2$CO$_2$ can possibly be folded into a nanotube form~\cite{X.Guo2016}.

The first-principles calculations for armchair and zigzag nanoribons of pristine and functionalized Ti$_2$C, 
Ti$_3$C$_2$, and V$_2$C~\cite{S.Zhao2015} have shown that i) there is a remarkable atomic
reconstruction at the edges of nanoribbons, ii) hydrogen functionalization can increase magnetic 
moment as compared with the pristine nanoribbon, and iii) armchair Ti$_2$CO$_2$ nanoribbon with a width 
of 7.34 \AA~exhibits a band gap of $\sim$1.0 eV, which is 
significantly larger than that of 2D Ti$_2$CO$_2$ ($\sim0.24$ eV)~\cite{M.Khazaei2013}. 
It is also shown that the hole and electron carrier mobility of the Ti$_2$CO$_2$ nanoribbon can be tuned by 
the width as well as the edge engineering~\cite{X.Zhang2015_2}. 
More comprehensive calculations reveal that the band gap of semiconducting MXene nanoribbons  
is determined by a combination of factors such as the quantum confinement, 
the energy location of edge states, and the strength of $d$-$d$ hybridization~\cite{L.Hong2016}.

Fabricating van der Waals heterostructures of MXenes, and their stacking with chalcogenides, graphene or 
other 2D systems can be an innovative route to develop materials with exotic properties. This is promising 
since similar complex systems have been 
experimentally realized~\cite{A.K.Geim2013,L.Britnell2012,L.A.Ponomarenko2011}. 
The basic idea behind the stacking of different 2D layers is the band gap engineering and combining the 
electronic properties of different 2D layers together into a single 3D crystalline composite with 
substantially tailored electronic properties. For example, 
the electronic structures of Graphene/Hf$_2$C(OH)$_2$~\cite{M.Khazaei2016_1}, 
MoS$_2$/Ti$_2$C, MoS$_2$/Ti$_2$CT$_2$ (T = F and OH)~\cite{L.Y.Gan2013_2}, 
silicine/Sc$_2$CF$_2$~\cite{H.Zhao2015}, 
Sc$_2$CF$_2$/Sc$_2$CO$_2$~\cite{Y.Lee2015}, 
and various transition metal dichalcogenides/M$_2$CO$_2$ (M=Sc, Ti, Zr, and HF)~\cite{Z.Ma2014,A.N.Gandi2017}  heterostructures 
have been studied based on first-principles calculations.

Among these studies, it has been found that the transition metal dichalcogenides/Sc$_2$CO$_2$ 
heterostructures show indirect band gaps in the range of 0.13--1.18 eV, which are suitable for electronic 
device applications~\cite{Z.Ma2014}. 
Figure~\ref{fig:hetero} shows the density of states, projected band structure, 
excess-depletion charge density ($\Delta\rho$), and electron localization function (ELF) for 
Graphene/Hf$_2$C(OH)$_2$ heterostructure~\cite{M.Khazaei2016_1}. These results clearly show that the 
graphene is $n$-doped by Hf$_2$C(OH)$_2$ and this is simply because of the ultra low work function of 
Hf$_2$C(OH)$_2$ 
($\sim2.32$ eV) as compared with graphene ($\sim4.26$ eV)~\cite{M.Khazaei2016_1}. 
However, the Dirac-cone like energy dispersion of the graphene remains intact 
[see Fig.~\ref{fig:hetero}(b)], and is simply shifted to lower energies at $-1.1$ eV by receiving electrons from 
Hf$_2$C(OH)$_2$. Figures~\ref{fig:hetero}(c) and \ref{fig:hetero}(d) show that electrons are mainly transferred 
from OH bonds in Hf$_2$C(OH)$_2$ to $\pi^\star$ states of the graphene. 
Since the graphene and Hf$_2$C(OH)$_2$ are 
semi-metallic and metallic, respectively, it is expected that the graphane/Hf$_2$C(OH)$_2$ heterostructure 
exhibits metallic current-voltage characteristics. 
Due to the small distance between hydrogen atoms and graphene (2.27 \AA), the electrons are localized 
either on Hf$_2$C(OH)$_2$ or on graphene. However, it is clear 
that the NFE state remains stable on the other side of Hf$_2$C(OH)$_2$ without graphene. 
Moreover, the ELF analysis indicates that there is no chemical bond between the graphene and 
Hf$_2$C(OH)$_2$~\cite{M.Khazaei2015}.


\section{Outlook}\label{sec:outlook}

The research on low-dimensional materials is still at a preliminarily stage and will stay as a very exciting 
field of science and technology for many years. In particular, the MXenes era has just emerged in the 
cutting-edge materials research arena. It has already been proved that MXenes are suitable materials for 
interdisciplinary research and collaboration between academia and industry. Synthesizing MXenes with 
a particular surface functionalization is one of the main experimental challenges in this field.
Since some of the MAX phases (Mo$_2$GaC) are low temperature superconductors~\cite{W.Jeitschko1983}, 
it is timely to examine the superconductivity properties of various MXenes. Our knowledge for photonic 
properties and solar harvesting applications of 
MXenes are still limited. It might be possible to grow lateral MXenes 
using chemical vapor deposition technique. Therefore, it might be interesting to study the lateral MXene 
heterostructures computationally.
It has been demonstrated theoretically that 2D topological insulators might be promising thermoelectric materials~\cite{Y.Xu2014}. 
Thus, it is worth examining the theormoelectric properties of the topological MXenes.
Catalytic and nanostructure properties of MXenes have not been well studied yet. 
There are still very few theoretical studies on electron transport, optical, and magnetic properties of MXenes.  
We hope that this review could provide a broad insight into the current status of research and applications 
of MXenes, and stimulate further studies.

\end{document}